\documentclass[fleqn,a4paper]{vch-book}
\usepackage[english]{babel}
\usepackage{graphicx}
\usepackage{amsbsy}
\usepackage{amsmath}
\usepackage{makeidx}
\usepackage{times}
\usepackage{tabularx}
\usepackage{booktabs}
\usepackage{amsfonts}
\usepackage{amssymb}
\renewcommand{\vec}[1]{\boldsymbol{#1}}

\begin{document}
\chapter{A First-Principles Scheme for Calculating the Electronic Structure of
Strongly Correlated Materials: GW+DMFT \tocauthor
{F. Aryasetiawan, S. Biermann, A. Georges}}

\label{chap:editbooks}

\authorafterheading{$^a$F. Aryasetiawan\footnote{Corresponding
    author}, $^b$S. Biermann, $^c$A. Georges} \affil{
$^a$Research Institute for Computational Sciences, AIST,
1-1-1 Umezono, Tsukuba Central 2, Ibaraki 305-8568, Japan,\\
$^b$Centre de Physique Theorique, Ecole Polytechnique,
91128 Palaiseau, France\\
$^c$Laboratoire de Physique Theorique
de l'Ecole Normale Superieure
24, rue Lhomond
75231 Paris Cedex 05
France}

\section{Introduction}

\label{preface}

The last few decades have witnessed a substantial progress in the field of
electronic structure of materials. Using density functional theory (DFT)
\cite{hohenberg,kohn} within the local density approximation (LDA) or
generalized gradient approximation (GGA) \cite{perdew} it is quite a routine
to calculate the electronic structure of relatively complicated materials
containing tens of atoms per unit cell. The success of LDA, however, is also
accompanied by a number of serious problems. It was noticed very early on that
when applied to calculate the band structures of s-p semiconductors and
insulators, it was found that the band gaps are systematically underestimated
by some tens percents. Apart from the too small gaps, the band dispersions are
very reasonable. This remarkable property of the LDA is still waiting for an
explanation since formally there is no theoretical justification for
identifying the one-particle Kohn-Sham eigenvalues as quasiparticle energies
observed in photoemission experiments. Applications to alkali metals also
indicate some problems, albeit less serious. When the band dispersions are
compared with photoemission data, they are found to be too wide by 10-30 \%.
Some many-body calculations of the electron gas \cite{yasuhara,maezono},
however, suggest that the band widths are actually widened compared with the
free-electron values and that the LDA performs better than it is commonly
believed. If this turned out to be true, photoemission data would presumably
need a complete revision. In any case, the LDA errors in s-p metals are
probably less significant compared to the band gap errors in semiconductors
and insulators.

A much more serious problem of the LDA arises when it is applied to calculate
the electronic structures of so-called ''strongly correlated systems''. We
have to be more precise with what we mean by correlations. Even in the
electron gas, correlation as conventionally defined is rather large. It is as
large as exchange that the two almost cancel each other leaving the
free-electron band essentially unchanged. Thus, it is more appropriate in our
case to define correlation as anything beyond the LDA rather than anything
beyond the Fock exchange since the former is usually our starting point in
electronic structure calculations of solids.

Strongly correlated systems are characterized by partially occupied localized
orbitals such as found in transition metal oxides or 4f metals. Here the
problem is often more of \emph{qualitative} rather than \emph{quantitative
}nature. It is often found that the LDA predicts a transition metal oxide to
be a metal whereas experimentally it is an antiferromagnetic insulator. To
cite some examples, LaMnO$_{3},$ famous for its colossal magnetoresistance,
and La$_{2}$CuO$_{4}$, a well-known parent compound of high-temperature
superconductors, are antiferromagnetic insulators but predicted to be metals
by the LDA \cite{pickett}. In cases where the LDA does predict the correct
structure, it is legitimate to ask if the one-particle spectrum is also
reproduced correctly. According to the currently accepted interpretation,
transition metal oxides may be classified as charge-transfer insulators
\cite{fujimori,sawatzky}, which are characterized by the presence of occupied
and unoccupied 3d bands with the oxygen 2p band in between. The gap is then
formed by the oxygen 2p and unoccupied 3d bands, unlike the gap in LDA, which
is formed by the 3d states (Mott-Hubbard gap). A\ more appropriate
interpretation is to say that the highest valence state is a charge-transfer
state: During photoemission a hole is created in the transition metal site but
due to the strong 3d Coulomb repulsion it is energetically more favourable for
the hole to hop to the oxygen site despite the cost in energy transfer. A
number of experimental data, notably 2p core photoemission resonance, suggest
that the charge-transfer picture is more appropriate to describe the
electronic structure of transition metal oxides. And of course in the case of
4f metals, the LDA, being a one-particle theory, is totally incapable of
yielding the incoherent part of the spectral function or satellite structures.

The difficulties encountered by the LDA discussed above have prompted a number
of attempts at improving the LDA. Notable among these is the GW approximation
(GWA), developed systematically by Hedin in the early sixties \cite{hedin}. He
showed that the self-energy can be formally expanded in powers of the screened
interaction $W$, the lowest term being $iGW,$ where $G$ is the Green function.
Due to computational difficulties, for a long time the applications of the GWA
were restricted to the electron gas. With the rapid progress in computer
power, applications to realistic materials eventually became possible about
two decades ago. Numerous applications to semiconductors and insulators reveal
that in most cases the GWA \cite{ferdi,aulbur} removes a large fraction of the
LDA band-gap error. Applications to alkalis show band narrowing from the LDA
values and account for more than half of the LDA error (although controversy
about this issue still remains \cite{ku}).

The success of the GWA in sp materials has prompted further applications to
more strongly correlated systems. For this type of materials the GWA has been
found to be less successful. For example, GW calculation on nickel
\cite{ferdini} does reproduce the photoemission quasiparticle band structure
rather well, as compared with the LDA one where the 3d band width is too large
by about 1 eV, but the too large LDA exchange splitting of 0.6 eV
(experimentally 0.3 eV) remains essentially unchanged. Moreover, the famous 6
eV satellite is not reproduced. Application to NiO \cite{ferdi-nio}, a
prototype of transition metal oxides, also reveals some shortcomings. One
problem is related to the starting Green's function, usually constructed from
the LDA Kohn-Sham orbitals and energies. In the LDA the band gap is very
small, about 0.2 eV compared with the 4 eV experimental band gap. A commonly
used procedure of performing a one-iteration GW calculation yields about 1 eV
gap, much too small. This problem is solved by performing a partial
self-consistency, where knowledge of the self-energy from the previous
iteration is used to construct a better starting one-particle Hamiltonian
\cite{ferdi-nio}. This procedure improves the band gap considerably to a
self-consistent value of 5.5 eV and at the same time increases the
LDA\ magnetic moment from 0.9 $\mu_{B}$ to about 1.6 $\mu_{B}$ much closer to
the experimental value of 1.8 $\mu_{B}$ . However, the GWA maintains the
Mott-Hubbard gap, i.e., the gap is formed by the 3d states as in the LDA,
instead of the charge-transfer gap. In other words, the top of the valence
band is dominated by the Ni 3d. A more recent calculation using a more refined
procedure of partial self-consistency has also confirmed these results
\cite{faleev}. The problem with the GWA appears to arise from inadequate
account of short-range correlations, probably not properly treated in the
random-phase approximation (RPA).

Attempts at improving the LDA to treat strongly correlated systems were
initiated by the LDA+U method
\cite{anisimov-ldau,anisimov0,sasha-ldau,anisimov}, which introduces on top of
the LDA Hamiltonian a Hubbard U term and a double-counting correction term,
usually applied to partially filled 3d or 4f shells. The LDA+U method is
essentially a Hartree-Fock approximation to the LDA+U Hamitonian. In the LDA,
the Kohn-Sham potential does not explicitly distinguish between occupied and
unoccupied orbitals so that they experience the same potential. In, for
example, transition metal oxides, where the 3d orbitals are partially
occupied, this leads to metallicity or to underestimation of the band gap. The
LDA+U cures this problem by approximately pushing down the occupied orbitals
by $U/2$ and pushing up the unoccupied orbitals by $U/2$, creating a lower and
upper Hubbard band, thus opening up a gap of the order of the Hubbard $U$. The
LDA+U method has been successfully applied to late transition metal oxides,
rare earth compounds such as CeSb, as well as to problems involving
metal-insulator transition and charge-orbital ordering.

More recently, the idea of the LDA+U was extended further by treating the
Hubbard U term in a more sophisticated fashion utilizing the dynamical
mean-field theory (DMFT) \cite{georges}. The DMFT is remarkably well suited
for treating systems with strong on-site correlations because the on-site
electronic Coulomb interactions are summed to all orders. This is achieved by
using a mapping onto a self-consistent quantum impurity problem, thereby
including the effects of the surrounding in a mean-field approximation. The
strength of the DMFT is its ability to properly describe Mott phenomenon or
the formation of local moments, which is the key to understanding many
physical properties in strongly correlated materials. The combination of LDA
and DMFT takes advantage of the first-principles nature of LDA while at the
same time incorporates local correlation effects not properly treated within
the LDA. The LDA+DMFT\ method \cite{anisimov2,anisimov1,sasha-dmft} has been
successfully applied to a number of systems by now.

In both the LDA+U and LDA+DMFT\ methods, two fundamental problems remain
unaddressed. First, the Hubbard U is usually treated as a parameter, and
second, the Hubbard U term contains interaction already included in the LDA
but it is not clear how to take into account this double-counting term in a
precise way. Thus, a truly first-principles theory for treating strongly
correlated systems is still lacking. In this article, we describe a dynamical
mean-field approach for calculating the electronic structure of strongly
correlated materials from first-principles\cite{silke,sun}. The DMFT is
combined with the GW method, which enables one to treat strong interaction
effects \cite{kotliar1}. One of the main features of the new scheme is that
the Hubbard U is calculated from first principles through a self-consistency
requirement on the on-site screened Coulomb interaction, analogous to the
self-consistency in the local Green's function in the DMFT. Since the GWA has
an explicit diagrammatic representation, the on-site contribution of the GW
self-energy can be readily identified and the scheme then allows for a precise
double-counting correction.

In the next two sections, we will give a summary of the GWA and DMFT,
describing their main features. In the fourth section we lay out the GW+DMFT
scheme, followed by a simplified application of the scheme to the excitation
spectrum of nickel. Finally we discuss some future challenges and directions.

\section{The \emph{GW }Approximation}

\subsection{Theory}

It can be shown that the self-energy may be expressed as \cite{hedin}%

\begin{equation}
\Sigma(1,2)=-i\int d3\;d4\;v(1,4)G(1,3)\frac{\delta G^{-1}(3,2)}{\delta
\phi(4)} \label{Sigma}%
\end{equation}
where $v$ is the bare Coulomb interaction, $G$ is the Green function and
$\phi$ is an external time-dependent probing field. We have used the
short-hand notation $1=(x_{1}t_{1})$. From the equation of motion of the Green function%

\begin{equation}
G^{-1}=i\frac{\partial}{\partial t}-H_{0}-\Sigma
\end{equation}%

\begin{equation}
H_{0}=h_{0}+\phi+V_{H}%
\end{equation}
$h_{0}$ is the kinetic energy and $V_{H}$ is the Hartree potential. We then
obtain
\begin{align}
\frac{\delta G^{-1}(3,2)}{\delta\phi(4)}  &  =-\delta(3-2)\left[
\delta(3-4)+\frac{\delta V_{H}(3)}{\delta\phi(4)}\right]  -\frac{\delta
\Sigma(3,2)}{\delta\phi(4)}\nonumber\\
&  =-\delta(3-2)\epsilon^{-1}(3,4)-\frac{\delta\Sigma(3,2)}{\delta\phi(4)}
\label{dG-1}%
\end{align}
where $\epsilon^{1}$ is the inverse dielectric matrix. The GWA is obtained by
neglecting the vertex correction $\delta\Sigma/\delta\phi$, which is the last
term in (\ref{dG-1}). This is just the random-phase approximation (RPA) for
$\epsilon^{-1}.$ This leads to%

\begin{equation}
\Sigma(1,2)=iG(1,2)W(1,2)
\end{equation}
where we have defined the screened Coulomb interaction $W$ by%

\begin{equation}
W(1,2)=\int d3v(1,3)\epsilon^{-1}(3,2)
\end{equation}
The RPA dielectric function is given by%

\begin{equation}
\epsilon=1-vP
\end{equation}
where%

\begin{align}
P(\mathbf{r,r}^{\prime};\omega)  &  =-2i\int\frac{d\omega^{\prime}}{2\pi
}G(\mathbf{r,r}^{\prime};\omega+\omega^{\prime})G(\mathbf{r}^{\prime
},\mathbf{r};\omega^{\prime})\nonumber\\
&  =2\sum_{i}^{occ}\sum_{j}^{unocc}\psi_{i}(\mathbf{r)}\psi_{i}^{\ast
}(\mathbf{r}^{\prime})\psi_{j}^{\ast}(\mathbf{r)}\psi_{j}(\mathbf{r}^{\prime
})\label{PRPA}\\
&  \times\left\{  \frac{1}{\omega-\varepsilon_{j}+\varepsilon_{i}+i\delta
}-\frac{1}{\omega+\varepsilon_{j}-\varepsilon_{i}-i\delta}\right\}
\end{align}
with the Green function constructed from a one-particle band structure
$\{\psi_{i},\varepsilon_{i}\}$. The factor of 2 arises from the sum over spin
variables. In frequency space, the self-energy in the GWA takes the form%

\begin{equation}
\Sigma(r,r^{\prime};\omega)=\frac{i}{2\pi}\int d\omega^{\prime}e^{i\eta
\omega^{\prime}}G(r,r^{\prime};\omega+\omega^{\prime})W(r,r^{\prime}%
;\omega^{\prime})
\end{equation}
We have so far described the zero temperature formalism. For finite
temperature we have%

\begin{equation}
P(\mathbf{r,r}^{\prime};i\nu_{n})=\frac{2}{\beta}\sum_{\omega_{k}%
}G(\mathbf{r,r}^{\prime};i\nu_{n}+i\omega_{k})G(\mathbf{r}^{\prime}%
,\mathbf{r};i\omega_{k})
\end{equation}%

\begin{equation}
\Sigma(r,r^{\prime};i\omega_{n})=-\frac{1}{\beta}\sum_{\nu_{k}}G(r,r^{\prime
};i\omega_{n}+i\nu_{k})W(r,r^{\prime};i\nu_{k})
\end{equation}

In the Green function language, the Fock exchange operator in the Hartree-Fock
approximation (HFA) can be written as $iGv$. We may therefore regard the GWA
as a generalization of the HFA, where the bare Coulomb interaction $v$ is
replaced by a screened interaction $W$. We may also think of the GWA as a
mapping to a polaron problem where the electrons are coupled to some bosonic
excitations (e.g., plasmons) and the parameters in this model are obtained
from first-principles calculations.

The replacement of $v$ by $W$ is an important step in solids where screening
effects are generally rather large relative to exchange, especially in metals.
For example, in the electron gas, within the GWA exchange and correlation are
approximately equal in magnitude, to a large extent cancelling each other,
modifying the free-electron dispersion slightly. But also in molecules,
accurate calculations of the excitation spectrum cannot neglect the effects of
correlations or screening. The GWA is physically sound because it is
qualitatively correct in some limiting cases \cite{hedin95}.

\subsection{The GW approximation in practice}

\begin{figure}
[ptb]
\begin{center}
\includegraphics[
height=2.3921in,
width=3.2526in
]%
{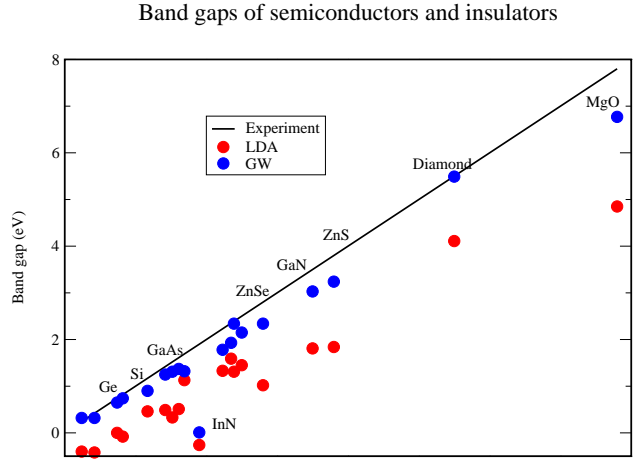}%
\caption{Band gaps of some selected semiconductors and insulators calculated
within the GWA compared with the LDA and experimental values. The GW data are
taken from \cite{kotani-gw}.}%
\label{bandgap}%
\end{center}
\end{figure}

The quality of the GWA may be seen in Figure (\ref{bandgap}), where a plot of
band gaps of a number of well known semiconductors and insulators is
displayed. It is clear from the plot that the LDA systematically
underestimates the band gaps and that the GWA substantially improves the
LDA\ band gaps. It has been found that for some materials, like MgO and InN,
significant error still remains within the GWA. The reason for the discrepancy
has not been understood well. One possible explanation is that the result of
the one-iteration GW calculation may depend on the starting one-particle band
structure. For example, in the case of InN, the starting LDA band structure
has no gap. This may produce a metal-like (over)screened interaction $W$ which
fails to open up a gap or yields too small gap in the GW calculation. A
similar behaviour is also found in the more extreme case of NiO, where a
one-iteration GW calculation only yields a gap of about 1 eV starting from an
LDA gap of 0.2 eV (the experimental gap is 4 eV) \cite{ferdi-nio,ferdi}.%

%TCIMACRO{\FRAME{ftbpFU}{3.2526in}{2.3921in}{0pt}{\Qcb{Band gaps of some
%selected semiconductors and insulators calculated within the GWA compared with
%the LDA and experimental values. The GW data are taken from \cite{kotani-gw}%
%.}}{\Qlb{bandgap}}{bandgap.eps}{\special{ language "Scientific Word";
%type "GRAPHIC";  maintain-aspect-ratio TRUE;  display "USEDEF";
%valid_file "F";  width 3.2526in;  height 2.3921in;  depth 0pt;
%original-width 9.3227in;  original-height 6.8424in;  cropleft "0";
%filename 'bandgap.eps';file-properties "XNPEU";}}}%
%BeginExpansion
%EndExpansion

The problems with the GWA arise when it is applied to strongly correlated
systems. Application to ferromagnetic nickel \cite{ferdini} illustrates some
of the difficulties with the GWA. Starting from the LDA band structure, a
one-iteration GW calculation does improve significantly the LDA band
structure. In particular it reduces the too large 3d band width bringing it
into much better agreement with photoemission data. However, the too large LDA
exchange splitting (0.6 eV compared with experimental value of 0.3 eV) remains
essentially unchanged. Moreover, the famous 6 eV satellite, which is of course
missing in the LDA, is not reproduced. These problems point to deficiencies in
the GWA in describing short-range correlations since we expect that both
exchange splitting and satellite structure are influenced by on-site
interactions. In the case of exchange splitting, long-range screening also
plays a role in reducing the HF value and the problem with the exchange
splitting indicates a lack of spin-dependent interaction in the GWA: In the
GWA the spin dependence only enters in $G$ but not in $W$.

Application to NiO, the prototype of Mott-Hubbard transition metal oxides,
reveals another difficulty with the one-iteration GWA. As already mentioned
previously, when the starting band structure is far from the experimental
quasiparticle band structure, a one-iteration GW calculation may not be
sufficient. This problem may be circumvented by performing a partial
self-consistent calculation in which the self-energy from the previous
iteration at a given energy, such as the Fermi energy of the centre of the
band of interest, is used to construct a new set of one-particle orbitals.
This procedure is continued to self-consistency such that the starting
one-particle band structure gives zero self-energy correction
\cite{ferdi-nio,ferdi,faleev}. A more serious problem, however, is describing
the charge-transfer character of the top of the valence band. The GWA
essentially still maintains the Mott-Hubbard band gap as in the LDA, i.e., the
top of the valence band is mainly of 3d character rather than the
charge-transfer character dominated by the 2p oxygen hole. As in nickel, the
problem with the satellite arises again. Depending on the starting band
structure, a satellite may be reproduced albeit at a too high energy. Thus
there is a strong need for improving the short-range correlations in the GWA
which may be achieved by using a suitable approach based on the dynamical
mean-field theory described in the next section.

\section{Dynamical Mean Field Theory}

Dynamical mean field theory (DMFT) \cite{georges} has originally been
developed within the context of \textit{models} for correlated fermions on a
lattice where it has proven very successful for determining the phase diagrams
or for calculations of excited states properties. It is a non-perturbative
method and as such appropriate for systems with any strength of the
interaction. In recent years, combinations of DMFT with band structure theory,
in particular Density functional theory with the local density approximation
(LDA) have emerged. The idea is to correct for shortcomings of DFT-LDA due to
strong Coulomb interactions and localization (or partial localization)
phenomena that cause effects very different from a homogeneous itinerant
behaviour. Such signatures of correlations are well-known in transition metal
oxides or f-electron systems but are also present in several elemental
transition metals.

The application of DMFT to \textit{real solids} relies on a
\textit{representability conjecture} assuming that local quantities such as
for example the local Green's function or self-energy of a solid can be
calculated from a local impurity model, that is one can find a
\textit{dynamical mean field} $\mathcal{G}_{0}$ and a Hubbard parameter $U$,
such that the Green's function calculated from the effective action
\begin{align}
S  &  =\int_{0}^{\beta}d\tau\sum_{m\sigma}c_{m\sigma}^{\dagger}(\tau
)\mathcal{G}_{0mm^{\prime}\sigma}^{-1}(\tau-\tau^{\prime})c_{m^{\prime}\sigma
}(\tau^{\prime})\nonumber\label{action}\\
&  +\frac{1}{2}\int_{0}^{\beta}d\tau\sum_{mm^{\prime}\sigma}U_{mm^{\prime}%
}n_{m\sigma}(\tau)n_{m^{\prime}-\sigma}(\tau)\nonumber\\
&  +\frac{1}{2}\int_{0}^{\beta}d\tau\sum_{m\neq m^{\prime}\sigma
}(U_{mm^{\prime}}-J_{mm^{\prime}})n_{m\sigma}(\tau)n_{m^{\prime}\sigma}(\tau)
\end{align}
coincides with the local Green's function of the solid. For a model of
correlated fermions on a lattice with infinite coordination number this
conjecture can rigorously be proven: it is a consequence of the absence of any
non-local contributions to the self-energy of the system. For a real solid the
situation is somewhat more complicated, not only due to the finite
coordination number but also to the difficulty to define the notion of
locality. This notion is crucial at both stages, for the construction of the
impurity model, where \textit{long-range} Coulomb interactions are mimicked by
\textit{local} Hubbard parameters\footnote{For a discussion of the
appropriateness of local Hubbard parameters see \cite{ferdi03}.},
%\cite{ferdi03},
%
%
%
%
%
%
and for the resolution of the model within DMFT, which approximates the full
self-energy of the model by a local quantity. Applications of DMFT to
electronic structure calculations (e.g. the LDA+DMFT method) are therefore
always defined within a specific basis set using localized basis functions.
Within an LMTO implementation for example locality can naturally be defined as
referring to the same muffin tin sphere. This amounts to defining matrix
elements $G_{L\mathbf{R},L^{\prime}\mathbf{R}^{\prime}}(i\omega)$ of the full
Green's function
\[
G(\mathbf{r},\mathbf{r}^{\prime},i\omega)=\sum_{LL^{\prime}\mathbf{R}%
\mathbf{R}^{\prime}}\chi_{L\mathbf{R}}^{\ast}(\mathbf{r})G_{L\mathbf{R}%
,L^{\prime}\mathbf{R}^{\prime}}(i\omega)\chi_{L^{\prime}\mathbf{R^{\prime}}%
}(\mathbf{r}^{\prime})
\]
and assuming that its local, that is ``on-sphere'' part equals the Green's
function of the local impurity model (\ref{action}). Here $\mathbf{R}%
,\mathbf{R}^{\prime}$ denote the coordinates of the centres of the muffin tin
spheres, while $\mathbf{r}$, $\mathbf{r}^{\prime}$ can take any values. The
index $L=(n,l,m)$ regroups all radial and angular quantum numbers. The
dynamical mean field $\mathcal{G}_{0}$ in (\ref{action}) has to be determined
in such a way that the Green's function $G_{impurityL,L^{\prime},\sigma}$ of
the impurity model Eq.(\ref{action}) coincides with $G_{L\mathbf{R},L^{\prime
}\mathbf{R}^{\prime}}(i\omega)$ if the impurity model self-energy is used as
an estimate for the true self-energy of the solid. This self-consistency
condition reads
\[
G_{impurity}(i\omega_{n})=\sum_{k}\left(  i\omega_{n}+\mu-H_{o}(k)-\Sigma
(i\omega_{n})\right)  ^{-1}%
\]
where $\Sigma,H_{0}$ and $G$ are matrices in orbital and spin space, and
$i\omega+\mu$ is a matrix proportional to the unit matrix in that space.

Together with (\ref{action}) this defines the DMFT equations that have to be
solved self-consistently. Note that the main approximation of DMFT is hidden
in the self-consistency condition where the local self-energy has been
promoted to the full lattice self-energy.

The representability assumption can actually be extended to other quantities
of a solid than its local Green's function and self-energy. In ``extended
DMFT'' \cite{si,kotliar,kajueter,sengupta} e.g. a two particle correlation
function is calculated and can then be used in order to represent the local
screened Coulomb interaction $W$ of the solid. This is the starting point of
the ``GW+DMFT'' scheme described in section 6.

\subsection{DMFT in practice}

Combinations of DFT-LDA and DMFT, so-called ``LDA+DMFT'' techniques
\cite{anisimov2}, have so far been applied to transition metals (Fe, Ni, Mn)
and their oxides (e.g. La/YTiO$_{3}$, V$_{2}$O$_{3}$, Sr/CaVO$_{3}$)
as well as elemental f-electron materials (Pu, Ce)
and their compounds.
In the most general formulation, one starts from a many-body Hamiltonian of
the form
\begin{align}
H &  =\sum_{\{im\sigma\}}(H_{im,i^{\prime}m^{\prime}}^{LDA}-H^{dc}%
)a_{im\sigma}^{+}a_{i^{\prime}m^{\prime}\sigma}\label{LDAUham}\\
&  +\frac{1}{2}\sum_{imm^{\prime}\sigma}U_{mm^{\prime}}^{i}n_{im\sigma
}n_{im^{\prime}-\sigma}\nonumber\\
&  +\frac{1}{2}\sum_{im\neq m^{\prime}\sigma}(U_{mm^{\prime}}^{i}%
-J_{mm^{\prime}}^{i})n_{im\sigma}n_{im^{\prime}\sigma},\nonumber
\end{align}
where $H^{LDA}$ is the effective Kohn-Sham-Hamiltonian derived from a
self-consistent DFT-LDA calculation. This one-particle Hamiltonian is then
corrected by Hubbard terms for direct and exchange interactions for the
``correlated'' orbitals, e.g. $d$ or $f$ orbitals. In order to avoid double
counting of the Coulomb interactions for these orbitals, a correction term
$H^{dc}$ is subtracted from the original LDA Hamiltonian. The resulting
Hamiltonian (\ref{LDAUham}) is then treated within dynamical mean field theory
by assuming that the many-body self-energy associated with the Hubbard
interaction terms can be calculated from a multi-band impurity model.

This general scheme can be simplified in specific cases, e.g. in systems with
a separation of the correlated bands from the ``uncorrelated'' ones, an
effective model of the correlated bands can be constructed; symmetries of the
crystal structure can be used to reduce the number of components of the
self-energy etc.

Despite the huge progress made in the understanding of the electronic
structure of correlated materials thanks to such LDA+DMFT schemes, certain
conceptual problems remain open: These are related to the choice of the
Hubbard interaction parameters and to the double counting corrections. An a
priori choice of which orbitals are treated as correlated and which orbitals
are left uncorrelated has to be made, and the values of U and J have to be
fixed. Attempts of calculating these parameters from constrained LDA
techniques are appealing in the sense that one can avoid introducing external
parameters to the theory, but suffer from the conceptual drawback in that
screening is taken into account in a static manner only \cite{ferdi03}.
Finally, the double counting terms are necessarily ill defined due to the
impossibility to single out in the LDA treatment contributions to the
interactions stemming from specific orbitals. These drawbacks of LDA+DMFT
provide a strong motivation to attempt the construction of an electronic
structure method for correlated materials beyond combinations of LDA and DMFT.

\section{GW+DMFT}

The idea behind combining the GWA and the DMFT is to take advantage of the
strong features of the two theories. The GWA, being based on RPA screening, is
capable of taking into account long-range correlations but does not describe
properly short-range correlations. On the other hand, the strength of the DMFT
is its ability to describe on-site correlations by means of a mapping of the
many-body problem onto an impurity problem where the on-site interactions are
summed to all orders. While the GWA is a fully first-principles theory, the
DMFT has been traditionally used in conjunction with a Hubbard model. Clearly,
it is desirable to combine the two theories into \ a consistent theory where
the parameters in the Hubbard model are determined self-consistently from
first-principles via the GWA.

Recently, first steps have been undertaken towards a combination of the GWA
and DMFT, both in a model context \cite{sun} and within the framework of
realistic electronic structure calculations \cite{silke}. The basic physical
idea of GW+DMFT is to separate the lattice into an on-site part and the rest.
The on-site self-energy is taken to be the impurity self-energy calculated by
the DMFT and the off-site self-energy is calculated by the GWA. Viewed from
the GWA, we replace the on-site GW self-energy by that of DMFT, correcting the
GW\ treatment of on-site correlations. Viewed from the DMFT, we add off-site
contributions to the self-energy approximated within the GWA, giving a
momentum dependent self-energy.

The impurity problem contains a Hubbard interaction U that is usually treated
as a parameter. In order to calculate the U, we introduce a self-consistency
condition that the U screened by the effective bath in the impurity model be
equal to the local projection of the global screened interaction W. This
condition is complementary to the condition imposed in the DMFT that the
impurity Green's function be equal to the local Green's function.

The above physical ideas can be formulated in a precise way using the
free-energy functional of Luttinger and Ward (LW). A generalization of the
original LW functional takes the form \cite{coa}(see also \cite{chitra})%

\begin{equation}
\Gamma(G,W)=Tr\ln G-Tr[(G_{H}^{-1}-G^{-1})G]-\frac{1}{2}Tr\ln W+\frac{1}%
{2}Tr[(v^{-1}-W^{-1})W]+\Psi\lbrack G,W] \label{LW}%
\end{equation}
$G_{H}^{-1}=i\omega_{n}+\mu+\nabla^{2}/2-V_{H}$ corresponds to the Hartree
Green's function with $V_{H}$ being the Hartree potential. The functional
$\Psi\lbrack G,W]$ is a generalization of the original LW $\Phi\lbrack G]$
functional, whose derivative with respect to $G$ gives the self-energy. A more
general derivation of (\ref{LW}) using a Hubbard-Stratonovich transformation
and a Legendre transformation with respect to both $G$ and $W$ may be found in
a later work \cite{chitra}. It is straightforward to verify that at
equilibrium the stationarity of $\Gamma$ yields%

\begin{equation}
\frac{\delta\Gamma}{\delta G}=0\rightarrow G^{-1}=G_{H}^{-1}-\Sigma
,\;\;\Sigma=\frac{\delta\Psi}{\delta G} \label{dGammadG}%
\end{equation}%

\begin{equation}
\frac{\delta\Gamma}{\delta W}=0\rightarrow W^{-1}=v^{-1}-P,\;\;P=-2\frac
{\delta\Psi}{\delta W}\label{dGammadW}%
\end{equation}
The functional $\Psi$ is divided into on-site and off-site components:%

\begin{equation}
\Psi=\Psi_{GW}^{off-site}[G^{RR^{\prime}},W^{RR^{\prime}}]+\Psi_{imp}%
^{on-site}[G^{RR},W^{RR}]\label{PsiGW+DMFT}%
\end{equation}
where $R$ denotes a lattice site. In the GWA the functional $\Psi$ is given by%

\begin{equation}
\Psi_{GW}[G,W]=\frac{1}{2}GWG \label{PsiGW}%
\end{equation}
The impurity part $\Psi_{imp}$ is generated from a local quantum impurity
problem defined on a single atomic site with an effective action%

\begin{align}
S  &  =\int d\tau d\tau^{\prime}\left[  -\sum c_{L}^{+}(\tau)\mathcal{G}%
_{LL^{\prime}}^{-1}(\tau-\tau^{\prime})c_{L^{\prime}}(\tau^{\prime})\right.
\nonumber\\
&  \left.  +\frac{1}{2}\sum:c_{L_{1}}^{+}(\tau)c_{L_{2}}(\tau):\mathcal{U}%
_{L_{1}L_{2}L_{3}L_{4}}(\tau-\tau^{\prime}):c_{L_{3}}^{+}(\tau^{\prime
})c_{L_{4}}(\tau^{\prime}):\right]  \label{Simp}%
\end{align}
The double dots denote normal ordering and $L$ refers to an orbital of angular
momentum $L$ on a given sphere where the impurity problem is defined. These
orbitals are usually partially filled localized 3d or 4f orbitals.

The GW+DMFT set of equations can now be readily derived from (\ref{LW}),
(\ref{PsiGW+DMFT}), and (\ref{PsiGW}) by taking functional derivatives of
$\Psi$ with respect to $G$ and $W$ as in (\ref{dGammadG}) and (\ref{dGammadW}):%

\begin{equation}
\Sigma=\Sigma_{GW}^{RR^{\prime}}(1-\delta_{RR^{\prime}})+\Sigma_{imp}%
^{RR}\delta_{RR^{\prime}} \label{SigGW+DMFT}%
\end{equation}%

\begin{equation}
P=P_{GW}^{RR^{\prime}}(1-\delta_{RR^{\prime}})+P_{imp}^{RR}\delta_{RR^{\prime
}} \label{PGW+DMFT}%
\end{equation}
In practice the self-energy is expanded in some basis set $\{\phi_{L}\}$
localized in a site. The polarization function on the other hand is expanded
in a set of two-particle basis functions $\{\phi_{L}\phi_{L^{\prime}}\}$
(product basis) since the polarization corresponds to a two-particle
propagator. For example, when using the linear muffin-tin orbital (LMTO)
band-structure method, the product basis consists of products of LMTO's. These
product functions are generally linearly dependent and a new set of optimized
product basis (OPB) \cite{ferdi} is constructed by forming linear combinations
of product functions, eliminating the linear dependencies. We denote the OPB
set by $B_{\alpha}=\sum_{LL^{\prime}}\phi_{L}\phi_{L^{\prime}}c_{LL^{\prime}%
}^{\alpha}$. To summarize, one-particle quantities like $G$ and $\Sigma$ are
expanded in $\{\phi_{L}\}$ whereas two-particle quantities such as $P$ and $W$
are expanded in the OPB set $\{B_{\alpha}\}$. It is important to note that the
number of $\{B_{\alpha}\}$ is generally smaller than the number of $\{\phi
_{L}\phi_{L^{\prime}}\}$ so that quantities expressed in $\{B_{\alpha}\}$ can
be expressed in $\{\phi_{L}\phi_{L^{\prime}}\}$, but not vice versa. In
momentum space, equations (\ref{SigGW+DMFT}) and (\ref{PGW+DMFT}) read%

\begin{equation}
\Sigma^{LL^{\prime}}(\mathbf{k},i\omega_{n})=\Sigma_{GW}^{LL^{\prime}%
}(\mathbf{k},i\omega_{n})-\sum_{\mathbf{k}}\Sigma_{GW}^{LL^{\prime}%
}(\mathbf{k},i\omega_{n})+\Sigma_{imp}^{LL^{\prime}}(i\omega_{n})
\label{Sig_a}%
\end{equation}%

\begin{equation}
P^{\alpha\beta}(\mathbf{k},i\omega_{n})=P_{GW}^{\alpha\beta}(\mathbf{k}%
,i\omega_{n})-\sum_{\mathbf{k}}P_{GW}^{\alpha\beta}(\mathbf{k},i\omega
_{n})+P_{imp}^{\alpha\beta}(i\omega_{n})\label{P_a}%
\end{equation}
The second terms in the above two equations remove the on-site contributions
of the GW self-energy and polarization, which are already included in
$\Sigma_{imp}$ and $P_{imp}$.

We are now in a position to outline the self-consistency loop which determines
$\mathcal{G}$ and $\mathcal{U}$ as well as the full $G$ and $W$ self-consistently.

\begin{itemize}
\item  The impurity problem (\ref{Simp}) is solved, for a given choice of
Weiss field $\mathcal{G}_{LL^{\prime}}$ and Hubbard interaction $\mathcal{U}%
_{\alpha\beta}$: the ``impurity'' Green's function $\ $%
\begin{equation}
G_{imp}^{LL^{\prime}}\equiv-\langle T_{\tau}c_{L}(\tau)c_{L^{\prime}}^{+}%
(\tau^{\prime})\rangle_{S}%
\end{equation}
is calculated, together with the impurity self-energy
\begin{equation}
\Sigma_{imp}\equiv\delta\Psi_{imp}/\delta G_{imp}=\mathcal{G}^{-1}%
-G_{imp}^{-1}.
\end{equation}
The two-particle correlation function%
\begin{equation}
\chi_{L_{1}L_{2}L_{3}L_{4}}=\langle:c_{L_{1}}^{\dagger}(\tau)c_{L_{2}}%
(\tau)::c_{L_{3}}^{\dagger}(\tau^{\prime})c_{L_{4}}(\tau^{\prime}):\rangle_{S}%
\end{equation}
must also be evaluated.

\item  The impurity effective interaction is constructed as follows:
\begin{equation}
W_{imp}^{\alpha\beta}=\mathcal{U}_{\alpha\beta}-\sum_{L_{1}\cdots L_{4}}%
\sum_{\gamma\delta}\mathcal{U}_{\alpha\gamma}O_{L_{1}L_{2}}^{\gamma}%
\chi_{L_{1}L_{2}L_{3}L_{4}}[O_{L_{3}L_{4}}^{\delta}]^{\ast}\mathcal{U}%
_{\delta\beta} \label{eff_inter}%
\end{equation}
Here all quantities are evaluated at the same frequency\footnote{Note that
$\chi_{L_{1}...L_{4}}$ does \textbf{not} denote the matrix element
$<L_{1}L_{2}|\chi|L_{3}L_{4}>$, but is rather defined by $\chi(r,r^{\prime
})=\sum_{L_{1}..L_{4}}\phi_{L_{1}}^{\ast}(r)\phi_{L_{2}}^{\ast}(r)\chi
_{L_{1}...L_{4}}\phi_{L_{3}}(r^{\prime})\phi_{L_{4}}(r^{\prime})$.} and
$O_{L_{1}L_{2}}^{\gamma}$ is the overlap matrix $\langle B_{\gamma}|\phi
_{L}\phi_{L^{\prime}}\rangle$. The polarization operator of the impurity
problem is then obtained as:%
\begin{equation}
P_{imp}\equiv-2\delta\Psi_{imp}/\delta W_{imp}=\mathcal{U}^{-1}-W_{imp}^{-1},
\end{equation}
where the matrix inversions are performed in the OPB $\{B_{\alpha}\}$.

\item  From Eqs.~(\ref{Sig_a}) and (\ref{P_a}) the full $\mathbf{k}$-dependent
Green's function $G(\mathbf{k},i\omega_{n})$ and effective interaction
$W(\mathbf{q},i\nu_{n})$ can be constructed. The self-consistency condition is
obtained, as in the usual DMFT context, by requiring that the on-site
components of these quantities coincide with $G_{imp}$ and $W_{imp}$. In
practice, this is done by computing the on-site quantities
\begin{align}
G_{loc}(i\omega_{n}) &  =\sum_{\mathbf{k}}[{G_{H}}^{-1}(\mathbf{k},i\omega
_{n})-\Sigma(\mathbf{k},i\omega_{n})]^{-1}\label{Glocal}\\
W_{loc}(i\nu_{n}) &  =\sum_{\mathbf{q}}[V_{\mathbf{q}}^{-1}-P(\mathbf{q}%
,i\nu_{n})]^{-1}\label{Wlocal}%
\end{align}
and using them to update the Weiss dynamical mean field $\mathcal{G}$ and the
impurity model interaction $\mathcal{U}$ according to:
\begin{align}
\mathcal{G}^{-1} &  =G_{loc}^{-1}+\Sigma_{imp}\label{update}\\
\mathcal{U}^{-1} &  =W_{loc}^{-1}+P_{imp}%
\end{align}
\end{itemize}

This cycle is iterated until self-consistency for $\mathcal{G}$ and
$\mathcal{U}$ is obtained (as well as on $G$, $W$, $\Sigma^{xc}$ and $P$).
When self-consistency is reached, $G_{imp}=G_{loc}$ and $W_{imp}=W_{loc}.$
This implies that at self-consistency, the second term in (\ref{Sig_a}) can be
rewritten as (in imaginary-time)%
\begin{equation}
\sum_{\mathbf{k}}\Sigma_{GW}^{LL^{\prime}}(\mathbf{k},\tau)=-\sum_{L_{1}%
L_{1}^{\prime}}W_{imp}^{LL_{1}L^{\prime}L_{1}^{\prime}}(\tau)G_{imp}%
^{L_{1}^{\prime}L_{1}}(\tau) \label{Sig_correction}%
\end{equation}

This shows that the local or on-site contribution of the GW self-energy is
precisely subtracted out, thus avoiding double counting. Eventually,
self-consistency over the local electronic density can also be implemented,
(in a similar way as in LDA+DMFT \cite{savrasov-pu,savrasov-functional}) by
recalculating $\rho(\vec{r})$ from the Green's function at the end of the
convergence cycle above, and constructing an updated Hartree potential. This
new density is used as an input of a new GW calculation, and convergence over
this external loop must be reached. While implementing self-consistency within
the GWA is known to yield unsatisfactory spectra \cite{holm}, we expect a more
favourable situation in the proposed GW$+$DMFT scheme since part of the
interaction effects are treated to all orders.

\subsection{Simplified implementation of GW+DMFT and application to
ferromagnetic nickel}

The full implementation of the proposed approach in a fully dynamical and
self-consistent manner is at the present stage computationally very demanding
and we regard it as a major challenge for future research. Here, we apply a
simplified scheme of the approach \cite{silke} to the electronic structure of
nickel in order to demonstrate its feasibility and potential. The main
simplifications made are:

\begin{enumerate}
\item  The DMFT local treatment is applied only to the $d$-orbitals, and we
replace the dynamical impurity problem by its static limit, solving the
impurity model (\ref{Simp}) for a frequency-independent $\mathcal{U}%
=\mathcal{U}(\omega=0)$.

\item  We perform a one-iteration GW calculation in the form \cite{ferdi}:
$\Sigma_{GW}=G_{LDA}\cdot W[G_{LDA}]$, from which the off-site part of the
self-energy is obtained.
\end{enumerate}

The local Green's function is taken to be
\begin{align}
G_{loc}^{\sigma}(i\omega_{n}) &  =\sum_{\mathbf{k}}[{G_{H}}^{-1}%
(\mathbf{k},i\omega_{n})-\Sigma_{GW}^{off-site}\label{Glocal2}\\
&  -(\Sigma_{imp,\sigma}-\frac{1}{2}Tr_{\sigma}\Sigma_{imp,\sigma}%
(0)+V_{xc}^{on-site})\,]^{-1}\nonumber
\end{align}
Thus, the off-site part is obtained from the GW self-energy whereas the
on-site part is derived from the impurity self-energy with a double-counting
correction of the form proposed in \cite{sasha}.

\begin{figure}
[ptb]
\begin{center}
\includegraphics[
height=2.1862in,
width=3.0874in
]%
{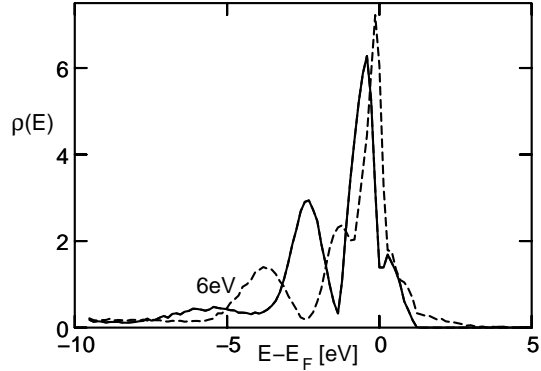}%
\caption{Partial density of states pf 3d orbitals of nickel (solid/dashed
lines give the majority/minority spin contribution) as obtained from the
combination of GW and DMFT \cite{silke}. For comparison with LDA and LDA+DMFT
results, see \cite{sasha}; for experimental spectra, see \cite{martensson}.}%
\label{dos}%
\end{center}
\end{figure}

\begin{figure}
[ptb]
\begin{center}
\includegraphics[
height=2.0928in,
width=2.9551in
]%
{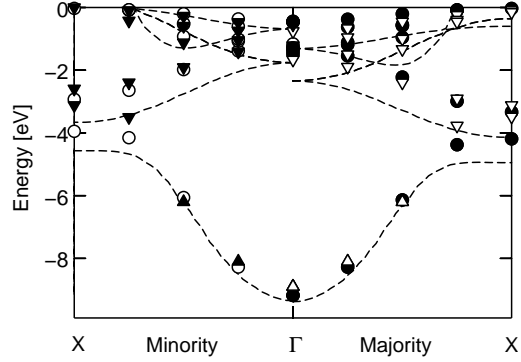}%
\caption{Band structure of nickel (majority and minority spins) from GW+DMFT
scheme \cite{silke} (circles) in comparison to the LDA band structure (dashed
lines) and experiments \cite{bunemann} (triangles down) and \cite{martensson}
(triangles up).}%
\label{bands}%
\end{center}
\end{figure}

We have performed finite temperature GW and LDA+DMFT calculations (within the
LMTO-ASA\cite{lmto} with 29 irreducible $\mathbf{k}$-points) for ferromagnetic
nickel (lattice constant 6.654 a.u.), using 4s4p3d4f states, at the Matsubara
frequencies $i\omega_{n}$ corresponding to $T=630K$, just below the Curie
temperature. The GW self-energy is calculated from a paramagnetic Green's
function, leaving the spin-dependence to the impurity self-energy. The
resulting self-energies are inserted into Eq.~(\ref{Glocal2}), which is then
used to calculate a new Weiss field according to (\ref{update}). The Green's
function $G_{loc}^{\sigma}(\tau)$ is recalculated from the impurity effective
action by QMC and analytically continued using the Maximum Entropy algorithm.
The resulting spectral function is plotted in Fig.(\ref{dos}). Comparison with
the LDA+DMFT results in \cite{sasha} shows that the good description of the
satellite structure, exchange splitting and band narrowing is indeed retained
within the (simplified) GW+DMFT scheme. We have also calculated the
quasiparticle band structure, from the poles of (\ref{Glocal2}), after
linearization of $\Sigma(\mathbf{k},i\omega_{n})$ around the Fermi level
\footnote{Note however that this linearization is no longer meaningful at
energies far away from the Fermi level. We therefore use the unrenormalized
value for the quasi-particle residue for the s-band ($Z_{s}=1$).}.
Fig.~(\ref{bands}) shows a comparison of GW+DMFT with the LDA and experimental
band structure. It is seen that GW+DMFT correctly yields the bandwidth
reduction compared to the (too large) LDA value and renormalizes the bands in
a ($\mathbf{k}$-dependent) manner.%

%TCIMACRO{\FRAME{ftbpFU}{3.0874in}{2.1862in}{0pt}{\Qcb{Partial density of
%states pf 3d orbitals of nickel (solid/dashed lines give the majority/minority
%spin contribution) as obtained from the combination of GW and DMFT
%\cite{silke}. For comparison with LDA and LDA+DMFT results, see \cite{sasha};
%for experimental spectra, see \cite{martensson}.}}{\Qlb{dos}}%
%{dos_gwdmft_0.0008.eps}{\special{ language "Scientific Word";
%type "GRAPHIC";  maintain-aspect-ratio TRUE;  display "USEDEF";
%valid_file "F";  width 3.0874in;  height 2.1862in;  depth 0pt;
%original-width 7.5031in;  original-height 5.3004in;  cropleft "0";
%filename '../Ni-GW+DMFT/dos_gwdmft_0.0008.eps';file-properties "XNPEU";}}}%
%BeginExpansion
%EndExpansion%

%TCIMACRO{\FRAME{ftbpFU}{2.9551in}{2.0928in}{0pt}{\Qcb{Band structure of nickel
%(majority and minority spins) from GW+DMFT scheme \cite{silke} (circles) in
%comparison to the LDA band structure (dashed lines) and experiments
%\cite{bunemann} (triangles down) and \cite{martensson} (triangles up).}%
%}{\Qlb{bands}}{gwdmft_xgx.eps}{\special{ language "Scientific Word";
%type "GRAPHIC";  maintain-aspect-ratio TRUE;  display "USEDEF";
%valid_file "F";  width 2.9551in;  height 2.0928in;  depth 0pt;
%original-width 7.5031in;  original-height 5.3004in;  cropleft "0";
%filename '../Ni-GW+DMFT/gwdmft_xgx.eps';file-properties "XNPEU";}}}%
%BeginExpansion
%EndExpansion

Because of the static approximation 3), we could not implement
self-consistency on $W_{loc}$ (Eq. (\ref{Wlocal})). We chose the value of
$\mathcal{U}(\omega=0)$ ($\simeq3.2eV$) by calculating the correlation
function $\chi$ and ensuring that Eq. (\ref{eff_inter}) is fulfilled at
$\omega=0$, given the GW value for $W_{loc}(\omega=0)$ ($\simeq2.2eV$ for
Nickel \cite{springer}).

\section{Conclusions}

The proposed GW+DMFT scheme avoids the conceptual problems inherent to
``LDA+DMFT'' methods, such as double counting corrections and the use of
Hubbard parameters assigned to correlated orbitals. The notion of
self-consistency condition on the on-site Green's function in the DMFT is
extended to the screened interaction. Analogous to the usual condition that
the impurity Green's function be equal to the on-site Green's function, we
demand that the on-site screened Hubbard $\mathcal{U}$\emph{ } of the impurity
be equal to the on-site projection of the global screened interaction
\emph{W}. In this fashion, the Hubbard $\mathcal{U}$ is determined from
first-principles. Since the GWA has a well-defined diagrammatic
interpretation, it is also possible to precisely take into account the
double-counting correction.

A number of issues are of immediate importance. Solving impurity models with
frequency-dependent interaction parameters \cite{motome,freericks,sun} is one
of the most urgent tasks as well as studies of various possible
self-consistency schemes. To study these aspects, we are now carrying out
GW+DMFT calculations on the electron gas. Applications to real materials are
both theoretically and computationally very challenging. Extension to the
two-particle Green's function is another field for future research
\cite{onida}.

\section{Acknowledgments}

This work was supported in part by NAREGI Nanoscience Project, Ministry of
Education, Culture, Sports, Science and Technology, Japan and by a grant of
supercomputing time at IDRIS Orsay, France (project number 031393).

\printindex
%\listoftables
%\listoffigures


\begin{thebibliography}{99}
\bibitem{coa}C.-O. Almbladh, U. von Barth and R. van Leeuwen, Int. J. Mod.
Phys. B \textbf{13}, 535 (1999)

\bibitem{lmto}O. K. Andersen, Phys. Rev. B \textbf{12}, 3060 (1975); O. K.
Andersen, T. Saha-Dasgupta, S. Erzhov, Bull. Mater. Sci. \textbf{26}, 19 (2003)

\bibitem{anisimov-ldau}V. I. Anisimov, J. Zaanen, and O. K. Andersen, Phys.
Rev. B \textbf{44}, 943 (1991)

\bibitem{anisimov0}V. I. Anisimov, I. V. Solovyev, M. A. Korotin, M. T.
Czyzyk, and G. A. Sawatzky, Phys. Rev. B \textbf{48}, 16929 (1993)

\bibitem{anisimov}For reviews, see V. I. Anisimov, F. Aryasetiawan, and A. I.
Lichtenstein, J. Phys.: Condens. Matter \textbf{9}, 767 (1997)

\bibitem{anisimov1}V. I. Anisimov et al.,        
J. Phys.: Condens. Matter \textbf{9}, 7359 (1997)

\bibitem{anisimov2}For reviews, see \emph{Strong Coulomb correlations in
electronic structure calculations}, edited by V. I. Anisimov, Advances in
Condensed Material Science (Gordon and Breach, New York, 2001)

\bibitem{ferdini}F. Aryasetiawan, Phys. Rev. B \textbf{46}, 13051 (1992)

\bibitem{ferdi-nio}F. Aryasetiawan and O. Gunnarsson, Phys. Rev. Lett.
\textbf{74}, 3221 (1995)

\bibitem{ferdi}F. Aryasetiawan and O. Gunnarsson, Rep. Prog. Phys.
\textbf{61}, 237 (1998)

\bibitem{ferdi03}F. Aryasetiawan et al., in preparation.

\bibitem{aulbur}W. G. Aulbur, L. J{\"o}nsson, and J. W. Wilkins, Solid State
Physics \textbf{54}, 1 (2000)

\bibitem{silke}S. Biermann, F. Aryasetiawan, and A. Georges, Phys. Rev. Lett.
\textbf{90}, 086402 (2003)

\bibitem{bunemann}J. B{\"u}nemann \emph{et al}, Europhys. Lett. \textbf{61},
667 (2003)

\bibitem{chitra}R. Chitra and G. Kotliar, Phys. Rev. B \textbf{63}, 115110 (2001)

\bibitem{faleev}S. V. Faleev, M. van Schilfgaarde, and T. Kotani, unpublished

\bibitem{freericks}J. K. Freericks, M. Jarrell and D. J. Scalapino, Phys. Rev.
B \textbf{48}, 6302 (1993)

\bibitem{fujimori}A. Fujimori, F. Minami, and S. Sugano, Phys. Rev. B
\textbf{29}, 5225 (1984)

\bibitem{georges}For reviews, see A. Georges, G. Kotliar, W. Krauth, and M. J.
Rosenberg, Rev. Mod. Phys. \textbf{68}, 13 (1996); T. Pruschke \emph{et al},
Adv. Phys. \textbf{44}, 187 (1995)

\bibitem{hedin}L. Hedin, Phys. Rev. \textbf{139}, A796 (1965); L. Hedin and S.
Lundqvist, \emph{Solid State Physics }vol. 23, eds. H. Ehrenreich, F. Seitz,
and D. Turnbull (Academic, New York, 1969)

\bibitem{hedin95}L. Hedin, Int. J. Quantum Chem. \textbf{54}, 445 (1995)

\bibitem{hohenberg}P. Hohenberg and W. Kohn, Phys. Rev. \textbf{136,} B864 (1964)

\bibitem{holm}B. Holm and U. von Barth, Phys. Rev. B \textbf{57}, 2108 (1998)

\bibitem{kajueter}H. Kajueter, Ph.D. thesis, Rutgers University, 1996

\bibitem{kohn}W. Kohn and L. J. Sham, Phys. Rev. \textbf{140}, A1133 (1965)

\bibitem{kotani-gw}T. Kotani and M. van Schilfgaarde, Solid State
Communications \textbf{121}, 461 (2002)

\bibitem{kotliar}G. Kotliar and H. Kajueter (unpublished)

\bibitem{kotliar1}For related ideas, see G. Kotliar and S. Savrasov, in
\emph{New Theoretical Approaches to Strongly Correlated Systems,} edited by A.
M. Tsvelik (Kluwer Academic Publishing, Dordrecht, 2001) (and the updated
version: cond-mat/0208241)

\bibitem{ku}W. Ku, A. G. Eguiluz, and E. W. Plummer, Phys. Rev. Lett.
\textbf{85}, 2410 (2000); H. Yasuhara, S. Yoshinaga, and M. Higuchi,
\emph{ibid}. \textbf{85}, 2411 (2000)

\bibitem{sasha-ldau}A. I. Lichtenstein, J. Zaanen, and V. I. Anisimov, Phys.
Rev. B \textbf{52}, R5467 (1995)

\bibitem{sasha-dmft}A. I. Lichtenstein and M. I. Katsnelson, Phys. Rev. B
\textbf{57}, 6884 (1998).

\bibitem{sasha}A. I. Lichtenstein, M. I. Katsnelson and G. Kotliar, Phys. Rev.
Lett. \textbf{87}, 067205 (2001)

\bibitem{maezono}R. Maezono, M. D. Towler, Y. Lee, and R. J. Needs, Phys. Rev.
B \textbf{68}, 165103 (2003)

\bibitem{motome}Y. Motome and G. Kotliar, Phys. Rev. B \textbf{62}, 12800 (2000)

\bibitem{martensson}H. M\aa rtensson and P. O. Nilsson, Phys. Rev. B
\textbf{30}, 3047 (1984)

\bibitem{northrup87}J. E. Northrup, M. S. Hybertsen, and S. G. Louie, Phys.
Rev. Lett. \textbf{59}, 819 (1987); \emph{ibid}. Phys. Rev. B \textbf{39},
8198 (1989)

\bibitem{onida}G. Onida, L. Reining and A. Rubio, Rev. Mod. Phys. \textbf{74},
601 (2002)

\bibitem{perdew}See, e.g., J. P. Perdew, K. Burke, and M. Ernzerhof, Phys.
Rev. Lett. \textbf{77}, 3865 (1996) and references therein.

\bibitem{pickett}See, e.g., W. E. Pickett, Rev. Mod. Phys. \textbf{62}, 433 (1989)

\bibitem{savrasov-pu}S. Savrasov and G. Kotliar, cond-mat/0106308

\bibitem{savrasov-functional}S. Savrasov, G. Kotliar and E. Abrahams, Nature
(London) \textbf{410}, 793 (2000)

\bibitem{sawatzky}G. A. Sawatzky and J. W. Allen, Phys. Rev. Lett. \textbf{53,
}2339 (1984)

\bibitem{sengupta}A. M. Sengupta and A. Georges, Phys. Rev. B \textbf{52},
10295 (1995)

\bibitem{si}Q. Si and J. L. Smith, Phys. Rev. Lett. \textbf{77}, 3391 (1996)

\bibitem{springer}M. Springer and F. Aryasetiawan, Phys. Rev. B \textbf{57},
4364 (1998)

\bibitem{sun}P. Sun and G. Kotliar, Phys. Rev. B \textbf{66}, 085120 (2002)

\bibitem{surh}M. P. Surh, J. E. Northrup, and S. G. Louie, Phys. Rev. B
\textbf{38}, 5976 (1988)

\bibitem{takada}Y. Takada, Phys. Rev. Lett. \textbf{87}, 226402 (2001)

\bibitem{tjeng}L. H. Tjeng, C. T. Chen, J. Ghijsen, P. Rudolf, and F. Sette,
Phys. Rev. Lett. \textbf{34}, 501 (1991)

\bibitem{yasuhara}H. Yasuhara, S. Yoshinaga and M. Higuchi, Phys. Rev. Lett.
\textbf{83}, 3250 (1999)
\end{thebibliography}
\end{document}